\documentclass[runningheads]{llncs}
\usepackage[T1]{fontenc}
\usepackage{graphicx}
\usepackage{hyperref}
\usepackage{graphicx} 
\usepackage{tikz}
\usetikzlibrary{positioning, shapes, arrows, calc, decorations.pathreplacing, arrows.meta, backgrounds, patterns, overlay-beamer-styles, matrix}
\usepackage{array}
\usepackage{tabularx}
\usepackage{pgfplots}
\pgfplotsset{compat=1.18}

\usepackage{color}

\usepackage{amsmath}

\begin{document}
%
\title{The Limits of Goal-Setting Theory in LLM-Driven Assessment}
%
%
\author{Mrityunjay Kumar\inst{1}\orcidID{0000-0003-2819-759X}}
\authorrunning{Mrityunjay Kumar}
%
\institute{Birla Institute of Technology and Science, Pilani, Pilani Campus, Vidya Vihar, Pilani, Rajasthan 333031, India \\
\email{mrityunjay.kumar@pilani.bits-pilani.ac.in}}




%
\maketitle              
\begin{abstract}
Many users interact with AI tools like ChatGPT using a mental model that treats the system as human-like—what we call Model H. According to goal-setting theory, increased specificity in goals should reduce performance variance. If Model H holds, then prompting a chatbot with more detailed instructions should lead to more consistent evaluation behaviour.

This paper tests that assumption through a controlled experiment in which ChatGPT evaluated 29 student submissions using four prompts with increasing specificity. We measured consistency using intra-rater reliability (Cohen’s Kappa) across repeated runs.

Contrary to expectations, performance did not improve consistently with increased prompt specificity, and performance variance remained largely unchanged. These findings challenge the assumption that LLMs behave like human evaluators and highlight the need for greater robustness and improved input integration in future model development.
\end{abstract}

\section{Introduction}

Artificial Intelligence (AI) has rapidly become mainstream technology. Consider these: 
\begin{itemize}
\item ChatGPT has almost 800 million active users in a week\footnote{\href{https://explodingtopics.com/blog/ai-statistics}{Exploding Topics - AI statistics}} 
\item It is predicted that within this decade, AI will improve employee productivity by 40\%\footnote{\href{https://www.pwc.com/us/en/tech-effect/ai-analytics/ai-predictions.html}{PwC - AI Predictions}}.
\item AI tool users will exceed 378M in 2025\footnote{\href{https://www.edge-ai-vision.com/2025/02/global-ai-adoption-to-surge-20-exceeding-378-million-users-in-2025/}{Statistia via EdgeAI + Vision Alliance}}.
\end{itemize}    

    
Thanks to the powerful conversational interfaces offered by OpenAI (ChatGPT), Anthropic (Claude), and Google (Gemini), a large number of non-tech, non-AI expert users use Generative AI actively. The top two uses for US AI users are in fairly generic tasks that everyone can engage in: asking questions and brainstorming\footnote{\href{https://www.aiprm.com/generative-ai-statistics/}{https://www.aiprm.com/generative-ai-statistics/}} 

We are interested in the users who do not understand the AI technology and are not in the technology domain (and hence understand AI much less). These users want to use AI to become more productive and mostly use tools with conversational interfaces (ChatGPT, Claude, DeepSeek, etc.). We name this User Persona \footnote{characterisation or representation of a typical user segment or end user - \href{https://en.wikipedia.org/wiki/Persona_(user_experience)}{Wikipedia}} NAIVE (\textbf{N}ot \textbf{AI} \textbf{V}eteran or \textbf{E}xpert). These NAIVE users may be naive in AI technology, but are experts in their own area.

In this paper, we want to examine the mental model they hold of these conversational AI systems they use. Mental models help humans make sense of their world \cite{norman2002}. When new technologies or tools are introduced, the user’s mental model of the system significantly influences how effectively the tool is accepted \cite{jungEffectsMentalModel2015}. 

There are many important questions related to the mental models that should be answered for this persona: 
\begin{itemize}
    \item What mental models do NAIVE users hold about the AI tool/system they use?
    \item Is this a useful mental model for them to have? Is it correct?
    \item How do they apply the mental model in generating their prompts?
    \item How does their mental model affect their productivity gains through AI?
\end{itemize}    

AI systems are technologies and tools as well, and we expect NAIVE users to hold a mental model that resembles a machine. However, the conversational nature of many Large Language Models (LLMs) may shift the mental model, and they may imbue the AI system with humanistic characteristics \cite{schneiderMentalModelShifts2025}. 

Therefore, we believe that there are two types of mental models that users may hold: 
\begin{itemize}
    \item Model M: AI system acts as a machine, and a prompt is a specification in a software sense - a program or instruction to the machine to command it to produce the desired output and conform to the specifications. 
    \item Model H: AI system acts as a human, and a prompt is a specification in a human sense - a goal or intent shared with the human to produce desired performance levels and achieve the goals.
\end{itemize}

Expert users are likely to have Model M as their mental model. For them, specific prompt strategies and prompt engineering practices help in detailing the specification and getting to detailed instructions, the 'full program'. 

NAIVE users are likely to have Model H as their mental model. For them, specific prompt strategies and prompt engineering practices help in creating detailed steps for the human to perform the actions to achieve the desired goals. 

For humans, goal specificity is an important attribute of goal-setting theory \cite{lockeBuildingPracticallyUseful2002}; it reduces variance in performance by reducing the ambiguity about what is to be attained \cite{lockeSeparatingEffectsGoal1989a}. As the goal specificity improves, the variance in performance reduces. 

Since Model H models an AI system as a human, the user is likely to assume that the system will demonstrate goal specificity. 

Does this hold for an AI system?  

Here are our research questions:
\begin{description}
    \item RQ1: How does the performance of the system depend on goal specificity?
    \item RQ2: How does the performance variance of the system depend on goal specificity?
\end{description}

We study these RQs in the context of a specific use case of an instructor (who holds Model H as their mental model) using a Chatbot (ChatGPT) to evaluate student responses to essay-type questions. We believe this is a domain that is very sensitive to the answer to these RQs and hence a very relevant use case. 

If goal specificity does not hold true for Model H, it may explain why NAIVE users are not able to produce good prompts \cite{zamfirescu-pereira_why_2023} - you need a good mental model to be able to produce good prompts.  

The paper reports the design, execution and findings of the experiment conducted to investigate these research questions. 

Section \ref{sec:related-work} discusses what the mental models and prompts literature says about the impact of mental models on prompts. Section \ref{sec:related-concepts} introduces the concepts we use in this study. Section \ref{sec:experimental-method} describes the experimental method used for this study and its context. Section \ref{sec:results} presents the results of the experiment, while Section \ref{sec:analysis-discussion} analyzes and discusses the findings. Section \ref{sec:conclusion-future} presents the conclusions and future work.

\section{Related Work}
\label{sec:related-work}
We want to understand how mental model impacts the prompts users create. 

Mental models are frameworks that individuals construct in order to support their predictions and understanding of the world around them \cite{johnson1983mental}. In the context of user interactions, Don Norman defines a mental model as a person’s internal representation of how a system works, grounded in their beliefs, observations, and inferences, which enables them to understand, predict, and mentally simulate the system’s behaviour \cite{norman2014some}. 

\textbf{What are prompts?}
A prompt is an input to a Generative AI model, that is used to guide its output \cite{schulhoff_prompt_2025}. They serve as the primary mechanism for translating user intentions into actionable outputs \cite{hewing_prompt_2024}. In the context of programming and software engineering, prompts are variously considered as specification, programs or instructions \cite{liang_prompts_2025,guy2024prompts}

While there are a lot of prompt heuristics available like In-Context Learning, Instruction-Following, Chain-of-Thought, Prompt tuning, Self-Consistency, Tree of Thoughts, etc., there isn't any unifying theory or clarity on what changes to prompts produce what kind of output changes \cite{kaddour_challenges_2023,chen_unleashing_2025}

"\textit{Merging prompt engineering and question asking into a unified field of study is essential for unlocking the full potential of human–machine collaboration}" \cite{sasson_lazovsky_art_2025}

Various discipline professionals are asked to learn prompt engineering, be in academic writing \cite{giray_prompt_2023}, medical profession \cite{mesko_prompt_2023,wang_prompt_2024} or higher education \cite{knoth_ai_2024}.

Goals to Intent mapping is a cognitive process, while Intent to Action mapping requires a good (LLM system) mental model, which many users do not possess \cite{subramonyam_bridging_2024}.

\textit{"Participants’ struggles with generating prompts, evaluating prompt effectiveness, and explaining prompt effects, primarily stemmed from ... models of system behaviour rooted in human-human interactions}." \cite{zamfirescu-pereira_why_2023}. 

Users are polite to machines too \cite{reeves1996media}, even experienced computer users apply social rules to their interaction with computers \cite{nass_computers_1994,zamfirescu-pereira_why_2023}

When new technologies or tools are introduced, the user’s mental model significantly influences how effectively the tool is accepted \cite{jung_effects_2015}

Chatbot users seem to start with a mental model of `Chatbot as Machine' but then transition to the model of 'Chatbot as Human' after the second prompt, "\textit{..[Users of chatbots] adopt conversational behaviours typical of human-to-human interaction, suggesting a cognitive transition in how they perceive and engage with AI systems.}" \cite{schneider_mental_2025,schneider_exploring_2024}.

"\textit{Most students showed signs of a human-to-human conversational structure in their prompting behavior. Students show a socially oriented communication style, which tends to be informal and focuses on sharing affective and emotional information}" \cite{knoth_ai_2024}.    

Humans expect more from the conversation when they think they are talking to humans than when they think they are talking to chatbots (Expectation violation) \cite{grimes_mental_2021}.

\section{Related Concepts}
\label{sec:related-concepts}
In this section, we will introduce a few concepts that are used in this study. As mentioned, we are studying the use case of an instructor who is evaluating student submissions (essay-type) using ChatGPT. It is important to interpret the terms 'Performance' and 'Performance variance' (as used in RQs) in the context of this use case. 

\textbf{Performance}

One of the ways to assess the performance of ChatGPT in an evaluation task is to compare its evaluation output with a human rater and use correctness as a measure. However, we use a more basic measurement. We compare the evaluations between two independent runs of the LLM with the same prompt and the same data. Given the stochastic nature of AI systems, these will not match perfectly, but we can use the level of agreement between the two evaluation runs as a measure of performance. Cohen's Kappa ($\kappa$)\footnote{\href{https://en.wikipedia.org/wiki/Cohen's_kappa}{Cohen's Kappa - Wikipedia}} measures this - it is an inter-rater reliability (IRR) metric. In this study, we use Kappa as a measure of the performance of the model (ChatGPT). Kappa varies between 0 (no agreement) to 1 (perfect agreement). The higher the Kappa, the better the performance. 

\textbf{Performance Variance}

Given Kappa as a measure of performance, performance variance can be checked by checking how Kappa values differ across multiple runs (keeping prompt and data the same). So if we have 6 runs, and hence 15 Kappas ($\binom{6}{2} = 15$), they represent performance values across the runs, and we can check their spread to see the variance. 

\textbf{Prompt Type vs. Goal Specificity}

Given that ChatGPT will always operate on a prompt, and a prompt consists of instructions as well as data, we use the term Prompt Type to represent a particular combination of instructions and data. A more specific goal in the context of our use case is a prompt containing more instructions and data (so that the evaluation task can be less ambiguous). In other words, goal specificity is represented as Prompt Type in this use case. 

\section{Experimental Method}
\label{sec:experimental-method}
This study investigates how the goal specificity (as represented by Prompt Types) influences the performance and performance variance (as represented by Kappa and its variance). 

\subsection{Context}
We designed our experiment around the use case of an instructor (the author) who has to grade the student submissions that are essay-type. These were submitted by the participants of an onboarding course module on Software Systems and Modelling run by the author at a SaaS product company. The submission was in response to a systems comprehension task that asked them to articulate their understanding of the given software system using a specified template\footnote{\href{https://bit.ly/c25template}{https://bit.ly/c25template}}. 
The software system used was Zoho Expense\footnote{\href{https://www.zoho.com/in/expense/}{https://www.zoho.com/in/expense/}}, and students were asked to explain their understanding of the product (after the instructor presented two use cases and gave a demo of the product showing these two use cases) in terms of (a) its components, (b) the behavior of each component, (c) interactions among components, and (d) the overall system behavior resulting from these interactions. 

\subsection{Evaluation Setup}

The implication of goal-setting theory is that with a more specific (includes more information and data) prompt, the performance should improve, and performance variance should decrease. We frame the behaviour of ChatGPT in evaluation tasks along two key dimensions derived from this theory.  
\begin{itemize}
    \item \textbf{Consistency}: Evaluation is consistent if performance (as measured by Kappa) improves when prompt specificity improves. 
    \item \textbf{Stability}: Evaluation is stable if the performance variance decreases when prompt specificity increases. 
\end{itemize}

These dimensions guide our research questions: RQ1 focuses on consistency, while RQ2 focuses on stability. Their operationalisation is described in Section~\ref{sec:metrics}.

To test the impact of prompt specificity, we defined four distinct Prompt Types. In addition to the basic details of the question asked and solution expected, different Prompt Types differed in the amount of additional details provided to help with the evaluation:

\begin{itemize}
    \item \textbf{Prompt Type 1 (P1)}: No rubric or model solution provided.
    \item \textbf{Prompt Type 2 (P2)}: Model solution provided, but no rubric.
    \item \textbf{Prompt Type 3 (P3)}: Rubric provided, but no model solution.
    \item \textbf{Prompt Type 4 (P4)}: Both rubric and model solution provided.
\end{itemize}

The rubric was generated using OpenAI's GPT-4 (o3) model, while the model solution was authored manually. All Prompt Types were used to evaluate the same set of 29 student submissions.

Each Prompt Type was used to run six independent evaluation sessions. Each run scored the student submissions using a Likert-style scale
from 1 (poor) to 5 (excellent) for each of the four evaluation dimensions (R1–R4)
as provided in the template. We selected R4 (‘overall system behaviour’) for this analysis, as it required integration of all previous dimensions and was thus a
good proxy for holistic understanding.

For each of the four Prompt Types, ChatGPT was run six times. This yielded $\binom{6}{2} = 15$ unique evaluation pairs per Prompt Type. Across all four Prompt Types, we obtained a total of 60 pairwise comparisons for analysis.

\subsection{Measuring Consistency and Stability}
\label{sec:metrics}

To evaluate the performance and variance, we computed Cohen’s Kappa ($\kappa$) between each pair of evaluation runs under a given Prompt Type. We used the weighted version of Kappa (with linear weights), which accounts for the ordinal nature of the 5-point scoring scale.

We interpret the results as follows:
\begin{itemize}
    \item \textbf{Mean Kappa} (average of all Kappa values) represents the performance of ChatGPT for a given Prompt Type. We check consistency by checking whether the Mean Kappa across the Prompt Types stay the same or differs. This can be assessed visually using boxplots and calculated statistically using one-way ANOVA. This addresses \textbf{RQ1}.
    \item \textbf{SD Kappa} (standard deviation of all Kappa values) represents the performance variance of the model for a given Prompt Type. We check stability by checking whether the SD Kappa across the Prompt Types stays the same or differs. This addresses \textbf{RQ2}.
\end{itemize}

We expect ChatGPT's grading behaviour to be consistent and stable, given our hypothesis of Model H.

\section{Results}
\label{sec:results}
\subsection{RQ1: How does the performance depend on goal specificity?}
To examine whether ChatGPT's grading is consistent, we computed weighted Cohen’s Kappa scores across repeated evaluations for each of the four Prompt Types. Each Prompt Type was used to evaluate a common set of student submissions across six independent ChatGPT runs, resulting in multiple pairwise comparisons (15 pairs per Prompt Type).

\begin{table}[ht]
\centering
\caption{Descriptive statistics of weighted Cohen’s Kappa values across prompt types.}
\label{tab:kappa_descriptive}
\renewcommand{\arraystretch}{1.2}
\begin{tabular}{>{\raggedright\arraybackslash}m{3.2cm}
                >{\centering\arraybackslash}m{2cm}
                >{\centering\arraybackslash}m{2cm}
                >{\centering\arraybackslash}m{2cm}
                >{\centering\arraybackslash}m{2cm}}
\hline
\textbf{Prompt Type} & \textbf{Description} & \textbf{Mean Kappa} & \textbf{SD} & \textbf{N (Pairs)} \\
\hline
1 & No rubric, no model       & 0.822 & 0.068 & 15 \\
2 & Model solution only       & 0.748 & 0.056 & 15 \\
3 & Rubric only               & 0.825 & 0.068 & 15 \\
4 & Rubric + model solution   & 0.765 & 0.073 & 15 \\
\hline
\end{tabular}
\end{table}

\begin{table}[ht]
\centering
\caption{ANOVA summary table comparing Kappa scores across prompt types.}
\label{tab:anova_kappa}
\renewcommand{\arraystretch}{1.2}
\begin{tabular}{>{\raggedright\arraybackslash}m{3.5cm}
                >{\centering\arraybackslash}m{1.2cm}
                >{\centering\arraybackslash}m{2.0cm}
                >{\centering\arraybackslash}m{2.0cm}
                >{\centering\arraybackslash}m{2.0cm}
                >{\centering\arraybackslash}m{2.0cm}}
\hline
\textbf{Source} & \textbf{Df} & \textbf{Sum Sq} & \textbf{Mean Sq} & \textbf{F value} & \textbf{Pr(>F)} \\
\hline
Prompt Type & 3 & 0.0693 & 0.0231 & 5.24 & 0.0029 \\
Residuals   & 56 & 0.2467 & 0.0044 &       &        \\
\hline
\end{tabular}
\end{table}

Descriptive statistics for each Prompt Type are shown in Table~\ref{tab:kappa_descriptive}. Prompt Type 3 (rubric only) yielded the highest Mean Kappa ($M = 0.825$), followed by Prompt Type 1 (no rubric or model; $M = 0.822$), Prompt Type 4 (rubric + model; $M = 0.765$), and Prompt Type 2 (model only; $M = 0.748$). Values are rounded to three decimal places, as shown in Table 1.

To test whether these differences were statistically significant, we performed a one-way ANOVA with Prompt Type as the independent variable and pairwise Kappa scores as the dependent variable. The results showed a significant main effect of Prompt Type, $F(3, 56) = 5.24$, $p = 0.0029$ (Table~\ref{tab:anova_kappa}).

We followed up with a Tukey HSD post-hoc test to identify which Prompt Types differed significantly. The test revealed that:

\begin{itemize}
    \item Prompt Type 2 produced significantly lower consistency than Prompt Type 1 ($p = 0.018$) and Prompt Type 3 ($p = 0.013$).
    \item Other comparisons (e.g., Prompt 4 vs. Prompt 3) were not statistically significant, though marginal differences were observed.
\end{itemize}
\begin{figure}
    \centering
    \includegraphics[width=0.9\linewidth]{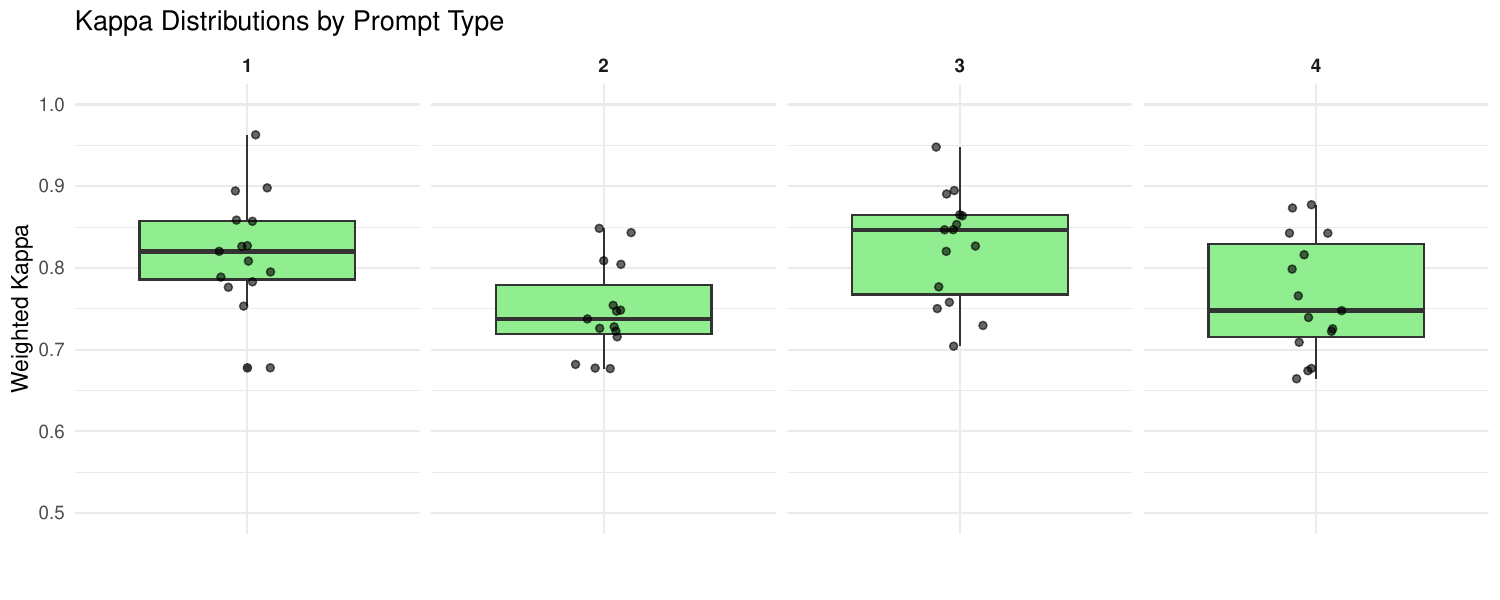}
    \caption{Boxplots of Weighted Kappa by Prompt Type}
    \label{fig:weighted-kappa-boxplot-byprompttype}
\end{figure}

\begin{figure}
    \centering
    \includegraphics[width=0.9\linewidth]{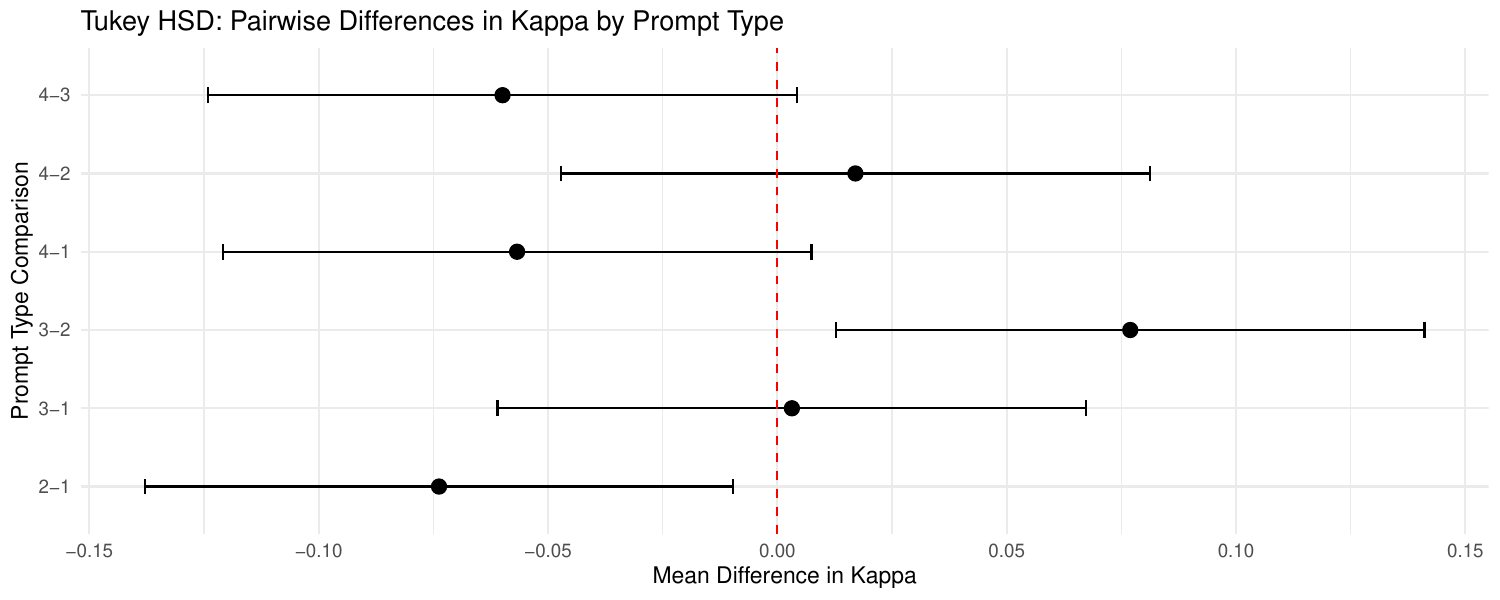}
    \caption{Tukey test - Pairwise mean differences}
    \label{fig:tukeyhsd-pairwise}
\end{figure}

These findings are visualised in Fig~\ref{fig:weighted-kappa-boxplot-byprompttype}, which shows the distribution of Kappa scores by Prompt Type, and Fig~\ref{fig:tukeyhsd-pairwise}, which plots the Tukey HSD pairwise differences and confidence intervals.

\subsection{RQ2: How does the performance variance depend on goal specificity?}

\begin{table}[ht]
\centering
\caption{Levene’s Test for Equality of Variances across Prompt Types}
\begin{tabular}{|>{\centering\arraybackslash}m{3cm}|>{\centering\arraybackslash}m{2cm}|>{\centering\arraybackslash}m{2cm}|>{\centering\arraybackslash}m{2.5cm}|>{\centering\arraybackslash}m{2.5cm}|}
\hline
\textbf{Test} & \textbf{DF Between} & \textbf{DF Within} & \textbf{F-value} & \textbf{p-value} \\
\hline
Levene’s Test & 3 & 56 & 0.498 & 0.685 \\
\hline
\end{tabular}
\label{tab:levene}
\end{table}

To address RQ2, we examined whether the variability in ChatGPT’s evaluation performance, measured through pairwise Cohen’s Kappa scores, differed across Prompt Types. For each Prompt Type, we computed 15 pairwise Kappa values from 6 independent evaluations and the standard deviations of these Kappa scores (Table~\ref{tab:kappa_descriptive}). 

To assess whether these differences in variability were statistically meaningful, we conducted Levene’s Test for homogeneity of variance \cite{howard_robust_1960}. The result, shown in Table~\ref{tab:levene}, was not significant ($F(3,56)=0.498$, $p=0.685$), indicating that the variability in consistency does not differ significantly across Prompt Types.

\section{Analysis and Discussion}
\label{sec:analysis-discussion}
\subsection{RQ1: How does the performance depend on goal specificity?}
The results of RQ1 show that ChatGPT’s performance is influenced by Prompt Type. Among the four types, Prompt Type 3 (rubric only) produced the best performance (as measured by Mean Kappa value), while Prompt Type 2 (model solution only) yielded the worst. Surprisingly, combining rubric and model (Prompt Type 4) did not outperform rubric alone, and introducing only a model solution (Prompt Type 2) led to the most inconsistent results. As Fig~\ref{fig:weighted-kappa-boxplot-byprompttype} also shows, the Mean Kappa values \textbf{do not} always increase when the prompt specificity improves. This contradicts the hypothesis that a more specific goal should improve performance. 

We derived RQ1 from goal-setting theory \cite{lockeSeparatingEffectsGoal1989a,lockeBuildingPracticallyUseful2002}, which predicts that increased specificity reduces performance variance. While this is well-established for humans, our findings show that not all forms of specificity benefit LLMs equally. Rubrics, structured and categorical, align well with how LLMs handle pattern-based reasoning. Model solutions, in contrast, may introduce ambiguity or distract the model with irrelevant surface features. As a result, users who assume LLMs behave like human graders (Model H) must be cautious when adding model examples or mixing input types.

More importantly, these results reveal a deeper limitation of current LLMs: ChatGPT performs worse in some cases when given more information, a counterintuitive and concerning outcome. This behaviour cannot be blamed on prompt wording or user error; it signals a structural weakness in how the model processes evaluative tasks. One explanation could be that rather than integrating multiple sources of guidance, the model appears to treat each prompt in isolation, often overfitting to surface cues. This fragility must be addressed in future LLM development by building mechanisms that allow stable integration of structured inputs like rubrics and examples.

Therefore, we can't assume that ChatGPT is consistent because, in some cases (as pointed out above), performance does not improve when prompt specificity improves.   

It is equally important to understand how the stability varies across Prompt Types. We now turn to RQ2, which investigates how performance variance depends on goal specificity.

\subsection{RQ2: How does the performance variance depend on goal specificity?}
The goal of RQ2 was to investigate whether the variability in performance differed across Prompt Types. Fig~\ref{fig:weighted-kappa-boxplot-byprompttype} shows the spread of the Kappas for each Prompt Type. All of them have similar spreads, even though the Interquartile Ranges for Prompt Types 1 and 2 are smaller than those of Prompt Types 3 and 4. 

While RQ1 showed significant differences in mean performance, the Levene’s test for homogeneity of variance \cite{howard_robust_1960} revealed no significant difference in the spread of Kappa scores across Prompt Types ($F(3,56)=0.498$, $p=0.685$). 

This result suggests that prompt specificity does not influence the variability of ChatGPT’s performance. All Prompt Types, regardless of the presence or absence of rubrics or model solutions, yielded a similar level of spread in performance. 

This points to an important distinction: while Prompt Types clearly affects the performance (even when it is not consistent), it does not appear to impact the performance variance. In other words, ChatGPT does not have performance stability.


This reinforces our earlier claim (from the discussion of RQ1) that Model H may not generalise to these systems.

This has implications for model design and deployment. Future model improvements should target not only accuracy and prompt responsiveness but also provide better stability. In high-stakes domains like assessment, where consistency and stability are critical, a good model must demonstrate both for them to be valuable for NAIVE users.

\section{Conclusion and future work}
\label{sec:conclusion-future}
This study was inspired by the goal-setting theory \cite{locke_building_2002} and investigated how the goal specificity (providing more information in the prompt, or less) affects the performance and performance variation of ChatGPT when evaluating student submissions. Through a controlled experiment using four types of prompts varying in the amount of information provided (by including or excluding rubric or model solution), we analysed both the consistency (RQ1) and stability (RQ2) across Prompt Types. 

Contrary to what the goal-setting theory suggests for humans (that providing more information improves the performance and reduces the variations in performance), our results show that adding more evaluative guidance, such as combining a rubric with a model solution, did not necessarily improve the performance or reduce the variation. This highlights a counterintuitive and concerning finding: ChatGPT can perform worse when given more relevant information, suggesting a structural limitation in its ability to integrate multiple sources of task guidance. We can say that it is neither consistent (more specificity increases performance) nor stable (more specificity reduces variance) in its performance. 

These findings challenge the assumption that large language models behave like human evaluators (Model H) and caution against direct applications of goal-setting theory to LLM-driven assessment tasks. More broadly, they call for greater attention to model robustness and input integration in future LLM development. For reliable use in educational contexts, it is not enough to make models accurate; they must also be consistent and stable across multiple runs with multiple prompt types. 

Future work will explore whether these effects hold across different domains, different types of additional information and question types, and how this can be corrected through fine-tuning or architecture-level improvements at the model level. 
\section{Acknowledgements}
\label{sec:acknowledgements}
We used OpenAI ChatGPT-4o (accessed between 20 June 2025 and 27 July 2025) for idea-generation, outlining, and language polishing. All outputs were reviewed and revised by the authors, who take full responsibility for the final content; ChatGPT is not listed as an author.

All data processing and figure generation were performed with \texttt{R}.

\bibliographystyle{splncs04}
\bibliography{prompts,reference}
%




\end{document}